\documentclass[useAMS,usenatbib]{mn2e}
\usepackage{epsfig}
\usepackage{graphicx}

\begin{document}
\voffset=-0.5 in

\title[Soft gamma repeaters and starforming galaxies]
{Soft gamma repeaters and starforming galaxies}

\author[S.B. Popov]{S.B. Popov$^{1}$ 
\thanks{E-mail: polar@sai.msu.ru}\\
$^1${\sl Sternberg Astronomical Institute, Universitetski pr. 13, 
Moscow 119992, Russia}  } 
\date{Accepted ......  Received ......; in original form ......
      }

\maketitle

\begin{abstract}
 We propose that the best sites to search for SGRs outside the Local group are
galaxies with active massive star formation. Different possibilities to
observe SGR activity from these sites are discussed. In particular we
searched
for giant flares from M82, M83, NGC 253, and NGC 4945 in the BATSE
data.  We present a list of potential candidates, however, in our opinion
no good candidates alike giant SGR flares were found. 
Still, hyperflares 
similar to the one of 27 December 2004 can be observed from
larger distances. From the BATSE data 
we select 5 candidates coincident with the galaxies Arp 299 and NGC
3256 which have very high rate of starformation,
and propose that they can be examples of hyperflares; however, this
result has low statistical significance.
\end{abstract}


\section{Introduction}

 Sources of 
soft gamma repeaters (SGRs) are one of the most puzzling types of neutron
stars. Now at least four of them are known in our Galaxy and in the
Large Magellanic Cloud
(we refer to \citet{wt2004} for a recent summary of
 all properties of SGRs).
 SGRs show three main types of bursts: 

\begin{itemize}
\item weak bursts, $L\la 10^{41}$~erg~s$^{-1}$;
\item intermediate bursts, $L\sim10^{41}$--$10^{43}$~erg~s$^{-1}$;
\item giant flares, $L\la 10^{45}$~erg~s$^{-1}$.
\end{itemize}

 Weak bursts are relatively frequent. About several hundreds were detected
from 4 sources during $\sim 25$ yrs, i.e. the average  rate is about 1 per
month per source. 
However, these bursts appear in groups during periods of activity of
a SGR. Duration is very short, $\sim 0.1$ s. 
 Intermediate bursts have typical durations $\sim$ few seconds and are much
more rare. These two types of bursts will not be discussed below.

Giant flares (GFs) are very rare. Only four GFs were observed (however, some
authors do not include the burst of SGR 1627-41 on June 18 1998 into this
list as it is slightly weaker then the others and shows a different pulse
structure).
These flares are extremely energetic. Typical duration of the initial spike
is about one second or smaller. The rate of GFs is very
uncertain as only four were detected, usualy it is assumed to be just 
(1/20 - 1/30) yrs$^{-1}$ per source. 
The latest GF detected on 27 December 2004 was suggested to be a
representative the  forth class of bursts -- ``supergiant flares''
or ``hyperflares'' (HFs). 
It exceeds previous GFs in the energy release
by two orders of magnitude, 
and below we consider this kind of events separately as they can be observed
from larger distances, and potentially can contribute to short-hard GRBs
detected by BATSE and other satellites (see \citet{ngpf2005, h2005} and
references therein).

 Being very interesting objects SGRs are very rare, 
probably due to a short life cycle, $10^4$ yrs, which can be
connected with a short duration of a magnetar activity. 
It would be very important to detect these sources outside the Local
group of galaxies to increase the sample.  
Especially it is interesting to understand the birth rate of SGRs and the
fraction of NSs which appear as these sources. Any solid data (even upper
limits) on the number of GFs and HFs detected from outside of the Local
group can help here a lot.

Here we want to discuss a possibility to observe SGRs outside the Local
group
of galaxies (for previous discussions of
extragalactic SGRs see \citet{d2001}  and 
recent e-prints by \citet{ngpf2005,h2005}).
Detection of such objects can give an opportunity to estimate
a fraction of NSs which appear as SGRs on larger statistics.
In this short note \footnote{This is just a brief discussion note, it is
submitted only to the ArXiv, and should be refered by its astro-ph number.
More elaborated results will be presented elsewhere (Popov, Stern in prep.).}
we mainly focus on regions of active starformation.  

We will discuss two approaches to find GFs and/or HFs from sources outside
the Local group:

\begin{itemize}
\item Close-by ($\la$10 Mpc)
galaxies with high starformation rate should give the main contribution
for detection of GFs. 
\item Few galaxies with extreme values of starformation rate 
(so-called ``supernova factories'') are the best
sights to search for rare HFs.
\end{itemize}


\section{Giant flares from local galaxies in the BATSE catalogue}

 As it is discussed by Heckman (1998)  inside 10 Mpc about 25\% 
of starformation is due to only four well known galaxies: M82, NGC 253,
 NGC 4945, and M83 (see Table 1). 
As probably BATSE could detect GFs 
similar to the prototype event of 5 March 1979 \citep{m1979}  
only from short distances ($\la 3$-$5$ Mpc), the
contribution of these four galaxies can be even more important.
The main idea which we put forward here is the following: 
close-by galaxies with high present day star formation rate are the best
sites to search for SGRs outside the Local group.

\begin{table}
 \centering
 \begin{minipage}{140mm}
  \caption{Close galaxies with high starformation rate}
  \begin{tabular}{@{}lrr@{}}
  \hline
   Name     & Distance, Mpc   &  SN rate per year \\
 \hline
 M 82       & 3.4             &       0.1-0.6    \\
 NGC 253    & 2.5             &       0.1-0.3    \\
 NGC 4945   & 3.7             &       0.1-0.5    \\
 M 83       & 3.7             &       0.1-0.5    \\
\hline
\end{tabular}
\end{minipage}
\end{table}

Supernova (SN) rates presented in the table are approximate ones as there
are no estimates better than a factor of 2-3 precision. 
\footnote{Note, that distances
are also uncertain, but the precision for them is much better.} 
We collected data from different papers.
For example, the rate for NGC253 is taken from \citet{e1998}
(see also \citet{p2001}).  
In the following we will use the conservative intermediate values: 0.4,
0.2, 0.2, and 0.1 for M82, NGC 253, NGC 4945,  and M83 correspondently.
In comparison with the Galactic rate of   
SN, these galaxies have significant enhancement (roughly 12, 6, 6, and 3
times correspondently), so we can expect proportionally higher number of
SGRs (and GFs from them). With the galactic rate $\sim $ 4 flares in 25
years, for BATSE (4.75 years equivalent of all-sky coverage) we can expect
roughly 8-9 GFs from M82, 4-5 GFs from each of NGC 253 and NGC 4945, and 
2 GFs from M83 (it total about 20 GFs from four galaxies during the BATSE
life cycle).        

It is useful to check if in the BATSE 
catalogue\footnote{http://cossc.gsfc.nasa.gov/batse/} there are potential
SGR-candidates in these four galaxies. We have to look for short bursts with
$T_{90}$ at least less than 2 seconds (the burst from SGR 1627-41 was longer
than initial strong spikes from three other SGRs). Another criterium is
fluence. Taking into account large distances to host galaxies 
we do not expect SGR candidates to be bright (expected fluences
$\la 10^{-7}$~erg~cm$^{-2}$). 
Then we have to select only 
relatively soft bursts, as GFs are softer than typical 
GRBs\footnote{A hard tail
in the spectrum of 5 March event starts at $E>430$~keV \citep{g1979}, so it
doesn't influence the third channel with $100<E<300$~keV; 
see also fig. 14.8 in \citep{wt2004}}.

All GRBs for which at least one of the four galaxies appears inside the
error box are given in Table 2. For each burst we give its number,
coordinates, error box radius, $T_{50}$ and $T_{90}$, maximum fluence
(maximum among four channels), and softness (the ratio of the 1st to the 3rd
channel). Coordinates and error box radii
are given in degrees.

\begin{table*}
 \centering   
 \begin{minipage}{140mm}
  \caption{GRBs coincident with SF galaxies}
  \begin{tabular}{@{}lrrrrrrr@{}}
  \hline
Trigger  &$\alpha$ & $\delta$ & Error & $T_{50}$,& $T_{90}$, & Maximum &
Softness\\
number   &         &          & box   &    s     &    s       & fluence &\\
&&&&&&& \\
\hline 
&&&&&&& \\
M82 &&&&&&& \\
 2054 & 164.33 & 66.15&  17.91&     ---  &     --- &      --- & --- \\
 2160 & 150.84 & 68.81&  2.15 &  11.072  &   123.136&     --- & ---\\
 2660 & 157.28 & 70.08 & 3.02 &   6.464  &    16.896 &    79.451e-08 & 0.613\\
 2821 & 124.84 & 60.59 & 13.53&  0.152   &    0.392 &     67.355e-09 & 0.122\\
 3118 & 117.57 & 80.37 & 34.15 &  0.136  &     0.232 &    18.628e-08 & 0.059\\
 3915 & 96.66 & 65.27 & 61.91  &  0.080    &   0.200   &  --- & --- \\
 6219 & 170.66 & 70.18 & 9.54  &  1.856  &     2.752    & 1.1506e-07 & 0.014\\
 6255 & 148.68&  60.79 & 12.71 &  ---       &  ---    &   ---  & --- \\
 6488 & 155.76&  76.41 & 9.94 &   1.152    &   2.240&     6.3462e-07 & 0.333\\
 6547 & 155.18&  62.23 & 13.58 &  0.029   &    0.097 &    1.0350e-07 & 0.129\\
 7297 & 140.07 & 76.39 & 9.53   & 0.438  &     1.141  &   6.5662e-07 & 0.050\\
 7552 & 137.56 & 65.9  & 6.15   & 11.648 &     63.296 &   5.8479e-07 & 1.122\\
 7970 & 136.87 & 64.49&  8.48   & 0.157 &      0.387   &  6.0876e-08 & 0.343\\
&&&&&&& \\
\hline
M83 &&&&&&& \\
 1485 & 202.01 &-29.98 & 9.65  &   ---     &  --- &     50.284e-08 & 0.501\\
 1510 & 198.84 &-34.35 & 7.29  &   ---     &  ---  &      --- & --- \\
 2349 & 203.15 &-31.88 & 6.19   &  1.696     &  4.032&   21.409e-08 & 0.426\\
 2384&  203.8 &-18.21  &17.81    & 0.128    &   0.192 &  84.657e-09 & 0.309\\
 2596&  211.51& -27.07 & 19.74    &---     &  ---      & --- & --- \\
 2756&  209.39& -23.16 & 9.65&     ---    &   ---      & --- & ---\\
 5444  &199.44 &-31.51 & 4.94 &    ---   &    ---      & --- & --- \\
 6447 & 191.44 &-36.6  &14.77  &   0.256     &  1.024  &  6.0229e-08 & 0.159\\
 6708&  206.73 &-29.7  &3.02    &   3.904   &   12.160  &  1.0452e-06 &0.051\\
 7361  &204.17 &-28.29 & 7.28    & 0.960   &    1.856   & 7.8501e-08 & 0.121\\
 7385 & 203.02 &-27.81 & 3.59     &---    &   ---       & --- & ---\\
 8076&  199.39 &-29.98 & 7.39     &0.075 &      0.218   & 3.1075e-07 & 0.072\\
&&&&&&& \\
\hline

NGC 253 &&&&&&& \\
2140  & 9.08 & -24.03&  4.34  & 6.784  &    19.072  &  48.766e-08 & 1.070 \\
2312  &14.72 &-33.56 & 8.93   &0.112  &     0.272  &  45.113e-08 &0.082\\
 2325 & 12.7 &-24.78 & 13.55  &  16.448&    22.528&    39.642e-08 & 0.218\\
 3908 & 11.56& -27.25&  7.96  & 5.344     & 13.920 &    59.805e-08& 0.203 \\
 6135 & 9.06 &-22.84 & 8.79   &1.856     &  4.032  &   1.4354e-06& 0.119\\
 6648 & 10.26& -26.97&  2.52  &24.512   &   88.384 &   1.7639e-06& 0.061\\
 6918 & 4.86 &-30.74 & 13.24  &---     &  ---    &  ---& ---\\
 7551 & 13.78& -24.45 & 3.75  &53.248 &    119.616&   6.8409e-07& 1.665 \\
&&&&&&& \\
\hline

NGC 4945 &&&&&&& \\
 108 & 201.31 &-45.41&  13.78     & 1.280&        3.136&  14.604e-07 & 0.259\\
 1167&  209.4 &-45.74&  10.04    &  7.552 &      24.576&  22.754e-08 & 0.330\\
 2405&  197.74& -44.32&  11.28  &   25.088 &     64.014&  83.646e-08 & 0.155\\
 2800 & 200.29& -47.94&  15.92 &     0.320  &     0.448&  35.153e-08 & 0.072\\
 3895 & 189.39& -47.72&  6.99 &      0.384   &    0.768&  42.483e-09 &0.135 \\
 6447 & 191.44& -36.6 & 14.77&       0.256    &   1.024&   6.0229e-08 &0.159\\

&&&&&&& \\
\hline
\end{tabular}
\end{minipage}
\end{table*}  

There are many short burst coincident with one of the four selected
galaxies.
(still, a significant number can be projected onto the region around a
particular galaxy by chance).
Only one of them -- 7970 -- is soft.
So, only this burst can be considered as a reasonable candidate. 
However, $T_{90}=0.387$~s seems to be long for a typical GF.

The main conclusion of this
analysis can be such 
that we found no good candidates GFs from hypothetical 
SGRs in M82, M83, NGC 253, and NGC 4945. 
However, if our assumption about softness of GFs can be soften several
candidates can be proposed.
This result can be used to put constraints on the number of SGR sources in
these galaxies, on the frequency of GFs or/and on the maximum flux during a
GF (Popov, Stern in prep.).

\begin{table*}
 \centering   
 \begin{minipage}{140mm}
  \caption{GRBs coincident with extreme SF galaxies}
  \begin{tabular}{@{}lrrrrrrr@{}}
  \hline
Trigger  &$\alpha$ & $\delta$ & Error & $T_{50}$,& $T_{90}$, & Maximum & \\
number   &         &          & box   &    s     &    s       & fluence \\
&&&&&& \\
\hline    
&&&&&& \\
Arp 299 &&&&&& \\
2265 & 180.2 & 59.57 & 8.32 &  0.256 &  0.456   &10.557e-08 \\
3118 & 117.57 & 80.37 & 34.15&    0.136 &  0.232  & 18.628e-08\\
6547 & 155.18 & 62.23 & 13.58 &  0.029 &  0.097 &  1.0350e-07\\
&&&&&& \\
\hline

NGC 3256 &&&&&& \\

2372 & 161.1& -36.01&  8.84&               0.072 & 0.256&
64.178e-09 \\
 2485 & 173.45 & -40.09 & 18.39&        0.128 & 0.176  & 34.750e-09
\\
\\
&&&&&& \\
\hline
\end{tabular}
\end{minipage}
\end{table*}  

\section{Hyperflares from Virgo and from galaxies with extreme starformation
rates}

  The situation with HFs (like the 27 Dec 2004 one) is quite 
different as the BATSE maximum detection distance
for such events is larger by about an 
 order of magnitude. 
 Having a limiting distance $\approx 50$ Mpc we can roughly estimate a
number of GFs and HFs from this volume. We assume that this number can be
scaled from the galactic one using the number of galaxies or
the starformation rate (SFR). 
Inside 50 Mpc we
can use several estimates. For example, \citet{d2001} uses the following
expression to obtain an estimate of a number of galaxies similar to the
Milky Way: $
N_{Gal}=0.0117 \,  h_{65}^3 \, R_{Mpc}^3. $  
For $R=50$ Mpc we obtain about 1500 galaxies. So, 
for 4.5 years of observation we can expect nearly 800 GFs
and about 200 HFs assuming 3 GFs and 1 HF observed in the Milky Way in 25 years.
Similar estimates can be obtained using estimates of \citet{bcw2004} and
\citet{gza1995}.
\citet{bcw2004} provide the following value for SFR density
at $z=0.1$: $0.01915\, M_{\odot}/{\rm yr}/{\rm Mpc}^3$. Inside 50 Mpc it
gives $\approx 10^4\, M_{\odot}/{\rm yr}/{\rm Mpc}^3$.
SFR for the Milky way is estimated to be few solar masses per year. 
So, the ratio is about few thousands. 
\citet{gza1995} estimate SFR in star-forming galaxies for $z\la0.045$ as
$0.013 \, M_{\odot}/{\rm yr}/{\rm Mpc}^3$. It gives $\approx 6800 \,
M_{\odot}/{\rm yr}/{\rm Mpc}^3$ inside 50 Mpc. All three estimates are in
good correspondence. So, having one HF in $\sim$30 years in our Galaxy we
can expect few hundreds HFs during the BATSE lifetime potentially detectable
by this satellite.

 The largest structure up to $R \sim 50$ Mpc is  the
Virgo cluster  of galaxies
(see \citet{bts1987} for all details about the cluster).
It includes about 1300 galaxies (including 130 spirals).
BATSE should be able to detect HFs from Virgo
cluster as fairly strong bursts. It is important to estimate an expected
number of GFs and HFs from Virgo. 
However, there are
several galaxies with significantly enhanced starformation. So, roughly we
can estimate that the SFR in Virgo is about few hundreds time larger than in
our galaxy. We can expect up to one hundred HFs during the BATSE
lifetime if we assume the rate in the Galaxy about 1 in  30-40 years.

Despite these optimistic predictions no anisotropy in distribution
of short GRBs or any correlations with known type of objects were found.
Here we want to adress another possibility --- observations of HFs
by BATSE from particular galaxies outside the Virgo cluster.
Of course, this large volume ($R\la 50$~Mpc) cannot be dominated by few
starforming galaxies, nevertheless, some peculiar objects can be considered
as important targets to search for HFs from extragalactic SGRs.

In the Universe there is a small amount of galaxies with extreme SFR --
``supernova factories''. They have core collapse SN rate up to two orders of
magnitude higher than in the Milky Way. 
Up to the limiting distance of detection of a HF by
BATSE there are two prominent objects of that kind: 
Arp 299 \citep{n2004} and NGC 3256 \citep{l2004}. 
We propose to look for HF candidates in the direction of these two galaxies.
\footnote{\citet{h2005} suggested that flares even stronger then the one of Dec 27 can
be expected from younger SGRs. Without any doubts the best place to look
for them are these two galaxies (and objects similar to them).}

Knowing a SN rate in a given galaxy we can obtain an estimate of a HF rate.
In the Galaxy only one HF in 30 years was
observed. (By the way it is roughly coincedent with the core collapse SN
rate.)
The SN rates in Arp 299 and NGC 3256 are about 1 per year. 
So, we can expect about 1 HF per year from each of them. 
During the BATSE lifetime few such events can be expected.

The fluence for the 27 Dec event was estimated in \citet{h2005} as 1.36 erg
cm$^{-2}$. The distance is about 15 kpc. For 40 Mpc (distance to Arp 299 and
NGC 3256) we can expect fluences about $10^{-7}$~erg~cm$^{-2}$.  

In the BATSE catalogue
\footnote{http://cossc.gsfc.nasa.gov/batse/} we looked for short
GRBs with known timing and fluences coincident with Arp 299 and NGC 3256.
On the whole (among 2704 GRB coordinates of which we used) 
error boxes of 12 appeared to
be coincident with Arp 299 and 6 with NGC 3256. From these set 
five short hard bursts with known fluences
were selected (Table~3): three from the direction of Arp 299 (for one
another very short burst -- the trigger number 3915 -- fluences are not given)
and two from NGC 3256 (however, another burst with the trigger number 6278 can
be a possible candidate).
All five GRBs look like short spikes in the BATSE data.
We propose that these 5 GRBs can be good candidates to be HFs.
Of course, some amount of events can be coincident with these galaxies
by chance, and such a probability is not low.\footnote{For example, three
bursts with large error boxes -- 3118, 3915 and 6547 -- appeared to be coincident
both with M82 and Arp 299; and 6447 with M83 and NGC 4945.} 
Still, we think that our finding is worth discussing.
Future and present day missions with better angular resolustion (like {\it
Swift} and, probably, {\it Integral} and {\it HETE})
can shed light on the association of short GRBs with galaxies with high SFR.


\section{Conclusions}

We discussed the connection between SGRs and starforming galaxies.
Our suggestion is that few well know galaxies with large SFR are the best
candidate sites to look for SGR flares. 
For the case of GFs we especially mention
such close-by galaxies as M82, M83, NGC 253, NGC 4945. 
For HFs we point to ``supernova factories'' Arp 299 and NGC 3256.

We found 5 candidates (3 in the direction of Arp 299 and 2 in the direction
of NGC 3256) which can be HFs.

\section*{Acknowledgments}

I want to thank Drs. B.E. Stern and M.E. Prokhorov
for many discussions and suggestions.
The work was supported by the RFBR grants 04-02-16720 and 03-02-16068.

\end{document}